%% file: main.tex
\documentclass[twocolumn]{aastex63}
\usepackage{inputenc}
\usepackage{comment}
\usepackage{colortbl}
\usepackage{graphicx}
\usepackage{soul}
\usepackage{hyperref}
\usepackage{multirow}
\usepackage{float}

\newcommand\Resolve{{\it XRISM}/Resolve }
\newcommand\FeKa{Fe K$\alpha$ }
\newcommand\FeKone{Fe K$\alpha_1$ }
\newcommand\FeKtwo{Fe K$\alpha_2$ }

\received{18 Jun 2026}
\revised{18 Aug 2026}
\accepted{19 Sep 2026}
\submitjournal{ApJ}

\shorttitle{X-ray Doppler}
\shortauthors{DiKerby \& Zhang 2026}
\graphicspath{{./}{figures/}}
\begin{document}

\title{First Detection of the Doppler Shift of X-ray Photons Due to a Telescope's Orbital Motion}

\author[0000-0003-2633-2196]{Stephen DiKerby}
\affiliation{Department of Physics and Astronomy \\
 Michigan State University, East Lansing, MI 48820, USA}

\author[0000-0002-2967-790X]{Shuo Zhang}
\affiliation{Department of Physics and Astronomy \\ Michigan State University, East Lansing, MI 48820, USA}
\begin{abstract}

We have detected for the first time the redshifting and blueshifting of X-ray photons due to the orbital motion of an X-ray telescope around the Earth. Current and planned X-ray telescopes like \Resolve and \textit{NewAthena} can achieve energy resolutions $R = E / \Delta E \approx 1000$, requiring spectroscopic calibration of physical effects such as the local standard of rest and the orbit of the Earth around the Sun. We construct an \textit{a priori} method for detecting a heretofore unexamined effect: the modulation of redshift due to the motion of an X-ray telescope around the Earth. We apply this pipeline to several \Resolve observations of narrow \FeKa emission lines from galactic center molecular clouds, and we detect evidence of orbital redshift modulation in the \FeKa emission at the $95\%$ confidence level. Our findings suggest that orbital redshift modulation is already present and detectable in X-ray data and motivates reexamination of how X-ray photon data is packaged in the emerging era of high-resolution X-ray spectroscopy.

\end{abstract}


\section{Introduction and Context}
\label{sec:Intro}

Until recently, X-ray telescopes \citep{Uhuru,HEAO2,Chandra} have had only limited energy resolution reaching $R = \frac{E}{\Delta E} \approx 400$. A new generation of instruments using calorimeter detectors such as Hitomi's Soft X-ray Spectrometer \citep{Hitomi} and the recently launched \Resolve \citep{2020arXiv200304962X,2024SPIE13093E..1PE} achieve $\approx 5\:\rm{eV}$ energy resolution in the $0.3-12\:\rm{keV}$ energy range ($R \approx 1000$). Future projects like the planned \textit{NewAthena} Integral Field Unit \citep{peille2025xrayintegralfieldunit} will match or exceed these improvements in X-ray spectroscopy, and X-ray detectors with energy resolution exceeding $R \geq 10^4$ may follow in the next few decades.

As the spectral resolution of a telescope improves, the range of systematic and physical effects that must be accounted for broadens. For example, modeling the \FeKa lines at $6.4 ~\rm{keV}$ with an instrument with $R = 100$ does not require accounting for broadening or redshifts with magnitudes far below $E/R \approx 64 \:\rm{eV}$. An instrument observing the same feature with $R=1000$ requires correction for finer effects like the earth's motion through the solar system and the solar system's movement with respect to the Local Standard of Rest \citep{2010MNRAS.403.1829S}. Corrections of this magnitude ($<10\:\rm{eV}$ or $<30 \:\rm{km/s}$) have been demonstrated in several \Resolve observations of astronomical objects with narrow spectral features \citep{2026ApJ...997L..20D,2026arXiv260327236P}.

The motion of an orbital telescope around the Earth is an effect observed and corrected with low-energy spectroscopy -- such as in Hubble with \texttt{DOPPCORR} \citep{2021cosd.book....5S} -- but never before considered for an X-ray telescope. The orbital motion of a telescope around the Earth imparts a $v = \pm \cos (i) \times 7.5 \:\rm{km/s}$ Doppler shift (where $i$ is the relative inclination of the satellites orbit with respect to the target, not with the equator) on incoming photons throughout a $\sim 90 \:\rm{minute}$ orbit \citep{1687pnpm.book.....N}. This effect is small and rapidly changes depending on the dot product of the telescope's instantaneous orbital motion with the unit vector towards the target of the observation. This effect may either appear as a coherent redshift in a spectral feature during a short time window, or as broadening if smeared out across an entire observation.

In the non-relativistic limit, the redshift imparted by the Doppler effect will change the energy of individual X-ray photons detected in a telescope by

\begin{equation} \label{eq:smallred}
    z \approx \frac{v}{c} \approx \frac{\Delta E}{E_0}
\end{equation}

\noindent where $v$ is the relative velocity of the satellite to the target, $E_0$ is the emitted photon energy, and $\Delta E$ is the change in photon energy. Even in high-resolution detectors like \Resolve the redshift imparted by orbital motion is small at the \FeKa lines: $\frac{v}{c} \times 6.4 ~\rm{keV} = 0.14 \:\rm{eV}$, far below the characterized energy resolution of \Resolve \citep{2025JATIS..11d2016P}. However, the statistics of many photons redshifted or blueshifted \textit{en masse} might be detectable. In our earlier work in \cite{2026ApJ...997L..20D} the redshift of the entire \FeKa complex of the galactic center molecular cloud G0.11-0.11 was constrained to $\sim 10 \:\rm{km/s}$, less than the expected $\approx 15 \:\rm{km/s}$ orbital redshift modulation. Furthermore, the XRISM gain recovery reports \footnote{available \hyperlink{https://heasarc.gsfc.nasa.gov/FTP/xrism/postlaunch/gainreports/}{for all XRISM observations}} reliably calculate the centroid of the calibration line at $5.9\:\rm{keV}$ to $\approx0.02 \:\rm{eV}$, corresponding to a velocity shift of $v = c \times \frac{\Delta E}{E} \approx 1 \:\rm{km/s}$, more than sufficient to detect the orbital redshift modulation. It is therefore the case that this effect might be detectable with current \Resolve observations, not just future X-ray instruments; the limitation is the degree to which redshift can be constrained for astrophysical sources, not just calibration lines.

Developing a methodology for identifying and filtering this effect can lead to even more precise measurements of X-ray spectral features, and should inform plans for future telescopes with even more precise energy resolution. For example, $\approx 0.1 \:\rm{eV}$ energy resolution at the \FeKa lines could necessitate photon-by-photon correction as the satellite orbits the Earth.

In this work, we create a pipeline to model orbital redshift in a generic X-ray telescope, and apply it to recent \Resolve observations of galactic center molecular clouds. In Section \ref{sec:ObsAna} we describe the steps of this pipeline, create a hypothesis test to determine the strength of evidence for this effect, and apply our pipeline blindly to four XRISM observations of strong \FeKa emission complexes. In Section \ref{sec:DiscConc} we discuss our results and interpret them in terms of high-resolution X-ray spectroscopy as a developing field.

\section{Data Pipeline and Analysis}
\label{sec:ObsAna}

\subsection{Telescope-Agnostic Pipeline}
\label{sec:Pipe}

Our approach to modeling orbital redshift in X-ray data is instrument-agnostic and uses as much as possible the typical conventions of X-ray spectral analysis. For illustrative purposes we use the XRISM observation $201052010$, an $\approx 120 \:\rm{ks}$ exposure of the galactic center molecular cloud G0.11-0.11 -- previously studied in \cite{2026ApJ...997L..20D} -- but analyses of individual \Resolve observations were conducted blindly and are discussed in greater detail in Section \ref{sec:XRIobs}. The overall approach is:

\textit{Using the on-off cadence detectable in a light curve, we construct time intervals that filter the total photon list based on whether the telescope is moving towards or away from the target, and test whether these opposing time intervals have different modeled redshifts.}

\begin{figure}
    \centering
    \includegraphics[width=\columnwidth]{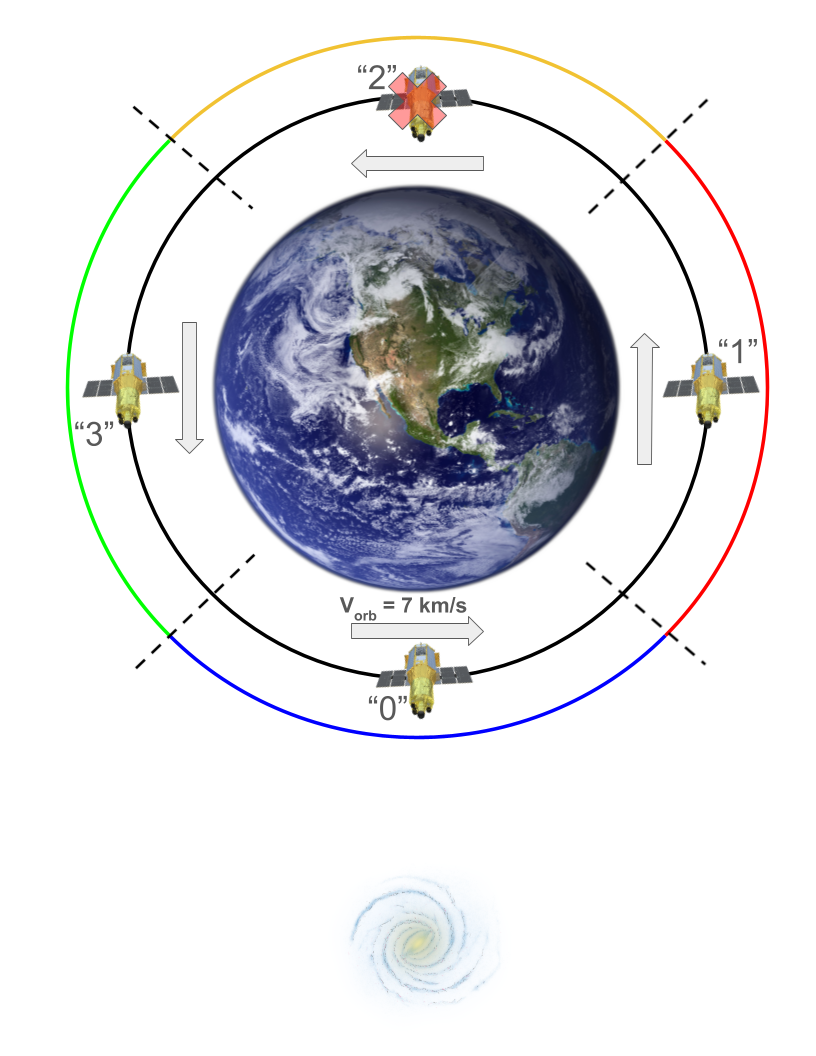}
    \caption{A not-to-scale schematic of the orbital motion of an X-ray telescope around the Earth with respect to its distant target (bottom). In our pipeline we divide the satellite orbit into four phases labeled in the diagram, centered on the period of inactivity due to its passage through the Earth's occultation of the target (red cross). The period of inactivity may not exactly align with the syzygy of the satellite, observations target, and the Earth.}
    \label{fig:OrbitalMotionDiagram}
\end{figure}

As shown in Figure \ref{fig:OrbitalMotionDiagram}, the orbital motion of a telescope around the Earth can be broadly divided into four quadrants, representing times during the observation when the telescope is mostly moving tangentially across the face of the Earth, away from the target, behind the Earth with a blocked view, and towards the target once emerging from behind the planet. For our purposes, we name these four quadrants phases 0, 1, 2, and 3, and color-code them consistently (blue, red, yellow, green shown in Figure \ref{fig:OrbitalMotionDiagram}).

\begin{figure}
    \centering
    \includegraphics[width=\columnwidth]{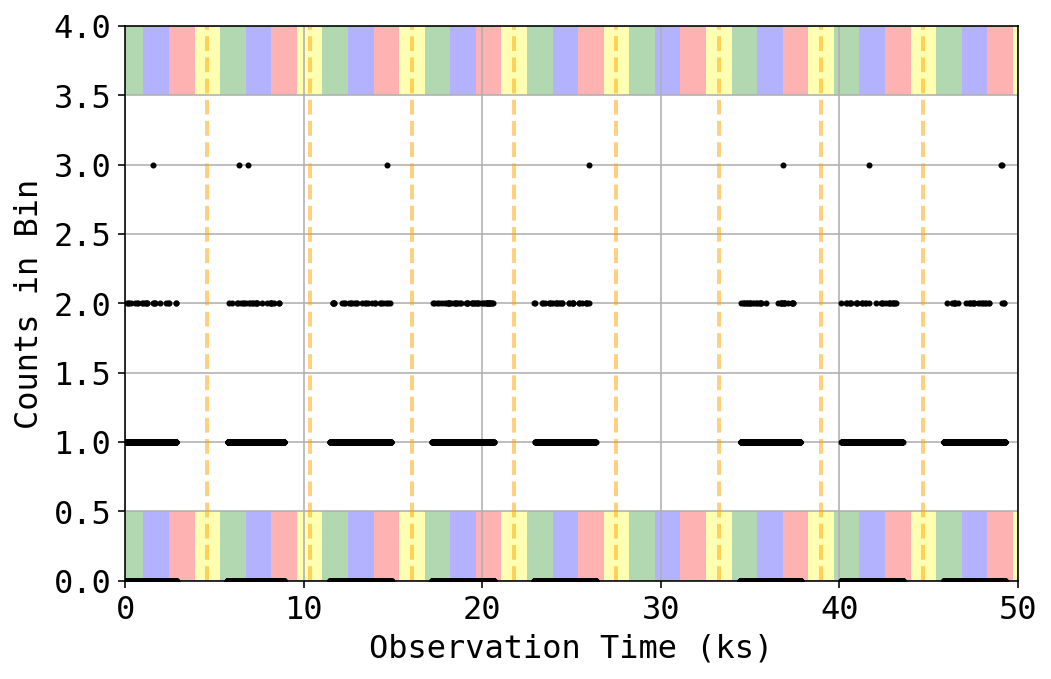}
    \caption{The first $50 \:\rm{ks}$ of the one-second-binned light curve of the XRISM observation $201052010$, showing gaps when the satellite ceases observing and the orbital timing solution with period $5728.5 \:\rm{s}$ (orange dashed lines) based on the periodicity of the gaps. The color-coded bars for the four phases of the light curve time filters are also plotted.}
    \label{fig:LightCurveSample}
\end{figure}

As usual, the X-ray photon data from the telescope is reprocessed, resulting in a clean event list, each photon characterized by its arrival direction, arrival time, and reconstructed energy. In our pipeline, the cleaned photon file is used to create a second-by-second light curve of the cleaned events via \texttt{xselect}.  The light curve in Figure \ref{fig:LightCurveSample} (using the first $50 \:\rm{ks}$ of the XRISM observation $201052010$ as an example) reveals a clear periodicity; once per orbit, the telescope passes behind the Earth and ceases observations of the target. This regular feature of the light curve can be exploited to divide the orbit of the telescope into intervals depending on the telescope's motion as shown in Figure \ref{fig:OrbitalMotionDiagram}.

In the light curve produced by \texttt{xselect}, the difference between the times of two adjacent bins reveals the rhythm of the telescope's orbit. Most bins will be separated from their neighbors by a single second, but when the telescope regularly passes behind the Earth or observations are otherwise interrupted the gap will be longer. Figure \ref{fig:LightCurveSample} includes these visible gaps between adjacent light curve bins for XRISM observation $201052010$, most of which have a duration almost exactly $2.28 \:\rm{ks}$.  We assume that the most regular and most common gaps are when the telescope passes behind the Earth, and that the interval between the start of these regular gaps (or where they would appear on a linear interpolation) is the orbital period of the telescope.

Indeed, for Obs ID 201052010 in Figure \ref{fig:LightCurveSample}, the periodicity of the gaps retrieves with great precision the orbital period of the satellite -- $5728 \:\rm{s}$ at the start of the observation, almost exactly in agreement with the orbital ephemerids\footnote{\hyperlink{https://www.n2yo.com/satellite/?s=57800}{https://www.n2yo.com/satellite/?s=57800}}.

With this tight constraint on the telescope's orbital period and the presumptive timing of its passages behind the Earth, we create composite time filters for each of the phases discussed above. We center phase 2 (when telescope is moving tangentially to the target behind the Earth) on the midpoint of each inactivity gap, presumably close to the midpoint of the target's occultation by the Earth. We divide the entire observation into quarter-orbit time slices from there. These time windows are displayed in corresponding colors in Figure \ref{fig:LightCurveSample}. Using these time windows, we use \texttt{xselect} to create phase-filtered event and spectrum files from the overall event list in each phase. All photons that arrived at the telescope when it is in the quadrant moving away from the target are grouped in the ``phase 1'' filtered spectrum, and so on.

X-ray spectral analysis proceeds normally for the phase-filtered spectra. RMFs, ARFs, and backgrounds can be generated or simulated. \texttt{xspec} can be used to analyze the phase-filtered spectra for some model with redshift as a parameter, obtaining a measurement of redshift $z$ for the phase 3 and phase 1 spectra. This can be conducted with any sufficiently strong and narrow feature in the X-ray spectrum. The difference between the phase 3 and 1 redshifts $\Delta v$ in $\rm{km/s}$ should encode the orbital motion of the telescope.

\subsection{Hypothesis Test}
\label{sec:Hypo}

As the orbital redshift has never before been observed in X-rays, we conduct a one-sided hypothesis test to evaluate the strength of evidence collected that the changes in observed redshift include periodic variations correlated with the satellite's orbit around the Earth. The \textit{null hypothesis}, that $\Delta v$ does not encode any discernible information about periodic redshift modulation, means that $\Delta v$ extracted identically from a series of observations would be distributed with population mean zero:

$\mathcal{H}_0$: The $\Delta v$ are drawn from a normal distribution with mean $\mu = 0 \:\rm{km/s}$

The alternate hypothesis is that $\Delta v$ captures some or all of the $15~\rm{km/s}$ orbital redshift expected during an orbit, modulo orbital inclination. Therefore the alternative hypothesis is:

$\mathcal{H}_1$: The $\Delta v$ are drawn from a normal distribution with mean $\mu > 0 \:\rm{km/s}$

\textit{A priori} we select the significance level of this hypothesis test at $\alpha = 0.05$, meaning that the null hypothesis of no period redshift modulation may be rejected at the 95\% confidence level if the p-value of the one-sided T test is less than 0.05.

\subsection{Application to XRISM Observations}
\label{sec:XRIobs}

We apply this pipeline to four XRISM observations of galactic center molecular clouds with narrow \FeKa lines: two of the Bridge cloud during the PV phase (Obs ID $300045010$, $300045020$) and one of Sgr C during Cycle 1 (Obs ID $201046010$), plus the aforementioned Obs ID $201052010$ of G0.11-0.11 examined in \cite{2026ApJ...997L..20D} and used in Figure \ref{fig:LightCurveSample}. Examining the gain reports of these four observations, we verify that no pixel has any failed gain solutions and that the calibration pixel intermittent solution appropriately tracks the continuous evolution. Because all these observations target diffuse sources larger than the \Resolve field of view, any biases in an individual pixel not accounted for in the XRISM data review or processing are unlikely to impact fitting results summed over the entire detector and duration.

All observations are processed identically and blindly according to the pipeline described above, with one manual step noted below. Using the spacecraft epherimedes of XRISM and the position of the galactic center at (RA, Dec) $ = 226.6, -28.9$, XRISM's motion encounters $14.7 \:\rm{km/s}$ of velocity change, close to the maximum possible. This makes the galactic center molecular clouds valuable targets for searching for redshift modulation compared to targets at other sky positions.

We use \texttt{heasoft v6.35} to reprocess the raw data with \texttt{xapipeline}. We exclude pixel-pixel coincident events, anomalous low-resolution secondary events, and periods of high particle background, and filter to only use high-resolution primary (Hp) events (always composing more than $98\%$ of the total events in any observation used herein). We exclude data from pixel 12, the calibration pixel, and pixel 27, which has unsuitable gain variation compared to the rest of the detector. As the molecular cloud targets of each observation are all of similar size to the field of view and spatial resolution of \Resolve, we extract the cleaned event list and one-second light curve of the entire remaining field of view.

Next, we use the generated light curves to produce a timing solution for each observation as in Figure \ref{fig:LightCurveSample}, using composite time intervals to generate filtered event and spectrum files for each phase. We group each spectrum to $S/N > 3$ in each bin using \texttt{ftgrouppha}. Because the width of the unbinned XRISM spectrum ($0.5 \:\rm{eV}$) and the energy resolution of XRISM at $6.4 \:\rm{keV}$ are greater than the energy shift imparted by the orbital motion, we do not expect this binning to impact our results. (We generate a comparison using unbinned spectra and the C-statistic \citep{1979ApJ...228..939C}, discussed in the Appendix, and come to the same conclusions as below). Because the vast majority of the emission we are analyzing is in the bright \FeKa lines, this binning primarily adjusts the bin width in the background continuum. Unsurprisingly but reassuringly, each ``phase 2'' filtered event file contains no photons, due to the occultation of the target by the Earth.

\begin{figure*}
    \centering
    \includegraphics[width=\linewidth]{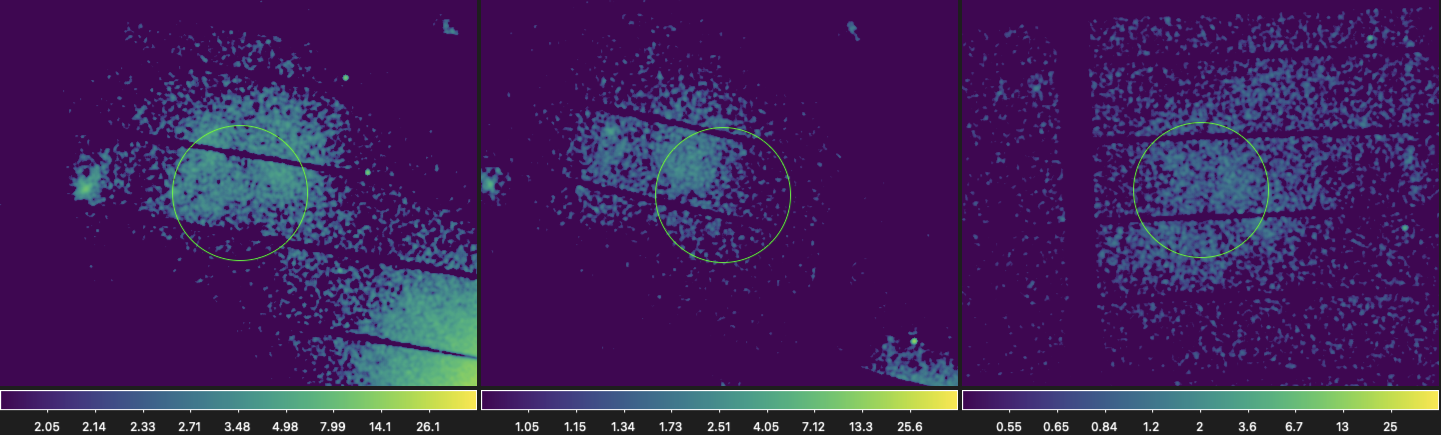}
    \caption{The $0.6-10.0 \:\rm{keV}$ \textit{XRISM}/Xtend fields used to create raytracing templates for (left to right) G0.11-0.11, the Bridge, and Sgr C in $4'$-diameter regions (green circle).}
    \label{fig:XtendTemplates}
\end{figure*}

We next use \texttt{rslmkrmf} to create a ``Large''-size RMF and \texttt{xaexpmap} to generate an exposure map for each phase-filtered spectrum. Creating simulated raytracing files for ARF generation is the only manual step of this process; we use the simultaneous \textit{XRISM}/Xtend $0.6-10.0 \:\rm{keV}$ image of each target in a $4'$-diameter cutout, shown for each target in Figure \ref{fig:XtendTemplates}, as template for this raytracing. Using these templates, we use \texttt{xaarfgen} to generate an ARF for each phase-filtered spectrum.  Finally, we generate a simulated background using the \texttt{rslnxbgen} command, as the target clouds are approximately the same size as the field of view of \Resolve and there is no clean background region.

\begin{figure}
    \centering
    \includegraphics[width=\columnwidth]{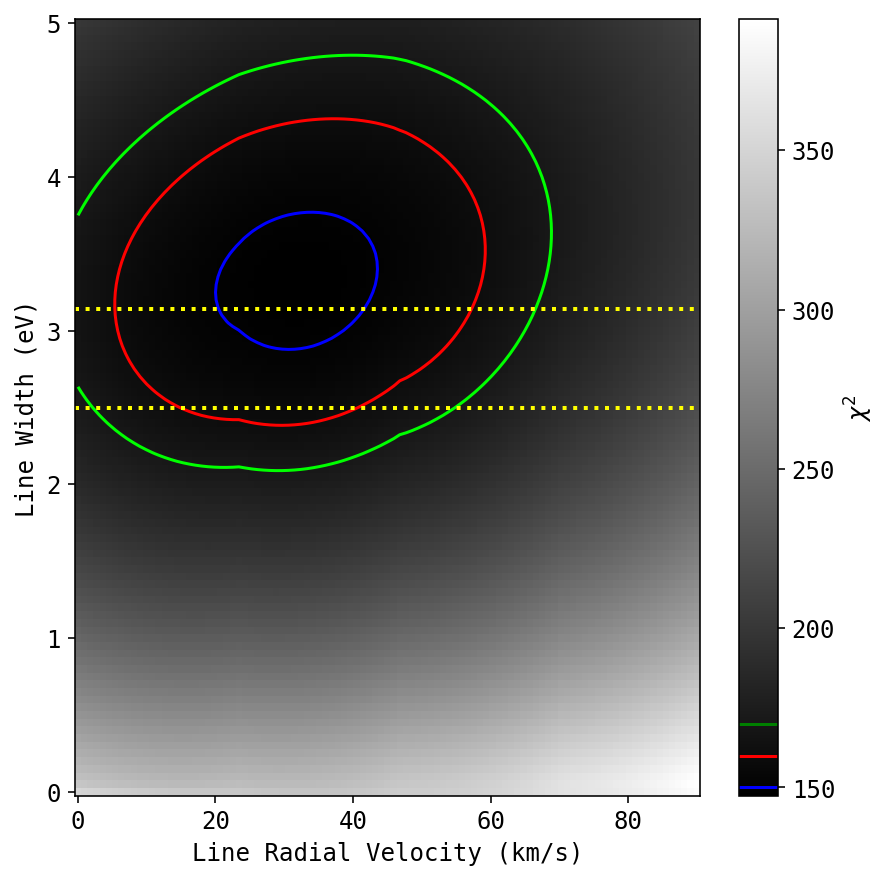}
    \caption{A $\chi^2$ contour plot for Obs ID $201052010$ comparing the goodness of fit for a grid of fitted velocity $v = cz$ and line widths. With no strong correlation in the contours ($\chi^2 = 150,160,170$ in blue, red, and green) of the two parameters, applying a fixed or fitted broadening to the \FeKa lines is not expected to impact the modeled redshift or its modulation. The \cite{1997PhRvA..56.4554H} \FeKone and \FeKtwo line widths are highlighted in yellow.}
    \label{fig:ZContour}
\end{figure}

\begin{figure}[h]
    \centering
    \includegraphics[width=\columnwidth]{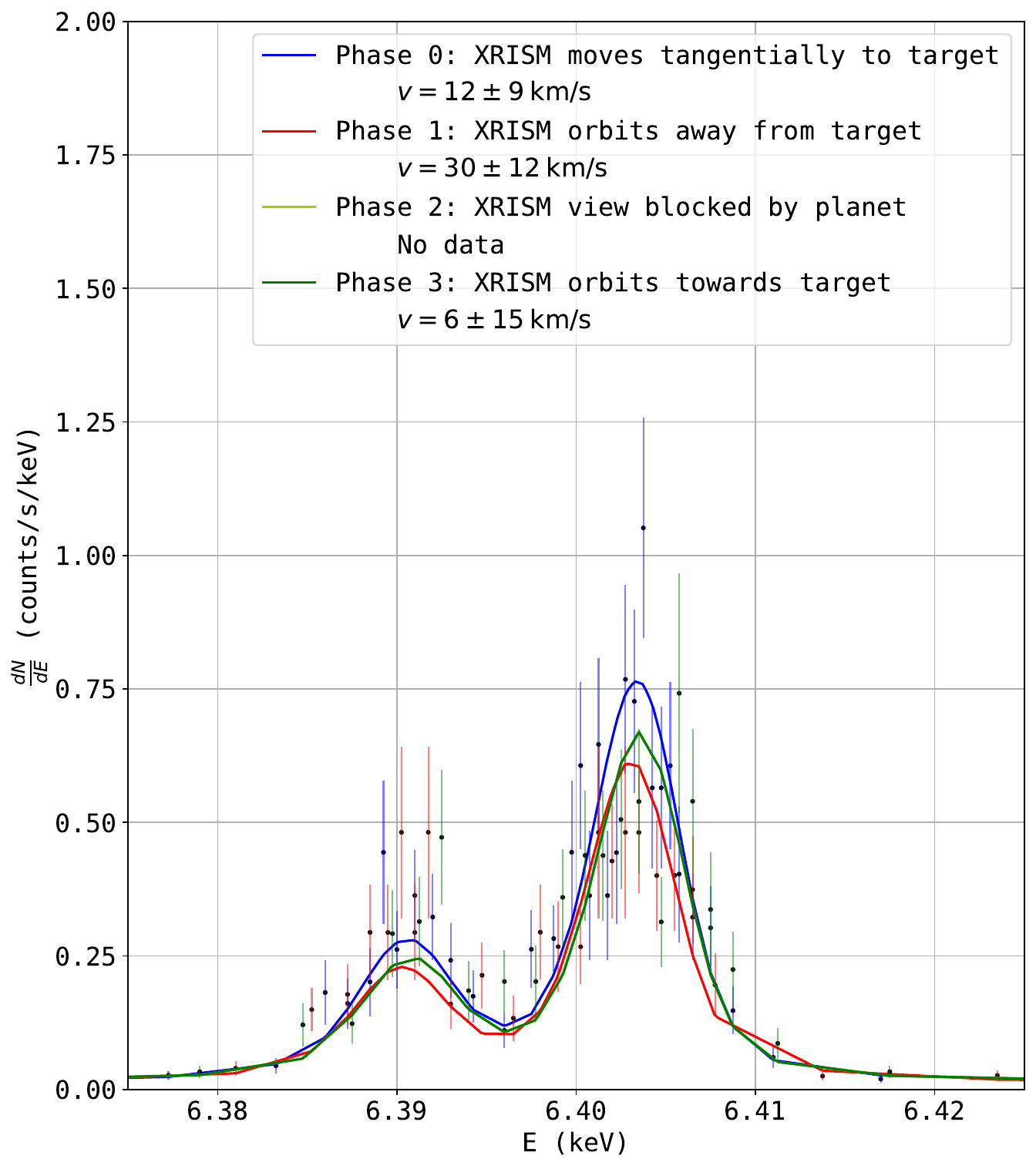}
    \caption{The phase-filtered spectra fitted with the \FeKa model from \cite{1997PhRvA..56.4554H} for Obs ID $201052010$, along with fitted redshift velocities. Phase 2 has no data due to occultation of the target by the Earth.}
    \label{fig:CompareSpec}
\end{figure}

We restrict \texttt{xspec} analysis to $6.0 ~\rm{keV} < E < 6.6 ~ \rm{keV}$, immediately around the very strong \FeKa line thought to be mostly produced by reflection of external X-rays \citep{2010ApJ...714..732P,2025A&A...695A..52S} and characterized at G0.11-0.11 in \cite{2026ApJ...997L..20D}. We fit the spectrum around the \FeKa complex with the model \texttt{ tbabs * (powerlaw + zfeklor)}. \texttt{tbabs} models interstellar absorption with cross-section data from \cite{Verner1996} and elemental abundance from \cite{Wilms2000}; we fix $n_H = 5 \times 10^{22} / \rm{cm^2}$, as $n_H$ is poorly constrained by this narrow energy band. \texttt{powerlaw} models the continuum emission from G0.11-0.11 and other sources appropriately for such a small energy range; we fix the slope of the power-law continuum $\Gamma = 0$ to account for the practically flat continuum in this small energy range.\footnote{After completing the hypothesis test, we revisited the fitting and verified that thawing these parameters or changing their values does not alter our results. Furthermore, our results do not change if the simulated point-source ARF provided for observation planning is used instead found \hyperlink{https://heasarc.gsfc.nasa.gov/docs/xrism/proposals/responses.html}{here}, or if no spectral background is loaded into \texttt{xspec}.}

\texttt{zfeklor} adapts the multi-Lorentzian model for \FeKa described in \cite{1997PhRvA..56.4554H} and fitted in detail in our previous work \citep{2026ApJ...997L..20D}. The only parameters of \texttt{zfeklor} are its normalization and critically the redshift $z$. In \cite{2026ApJ...997L..20D} we found no substantial velocity broadening in the \FeKa lines, and once again leaving broadening as a free parameter results in the line width equaling the quantum mechanical minima reported in \cite{1997PhRvA..56.4554H}. To check that our assumption of no broadening does not impact our results, we produce contour plots for redshift and line broadening for each observations, with the plot for the entire Obs ID $201052010$ in Figure \ref{fig:ZContour}. The lack of strong correlation between the contours of these two parameters (increasing line width within uncertainty does not dramatically change fitted redshift), and the fact that any biases would be applied equally to the phase 1 and 3 spectra, suggests that any redshift modulation would be detectable regardless of small changes in assumed line width.

\input{FitResults}

Fitting this model on the phase 0, 1, and 3 spectra, we find the best-fit parameters listed in Table \ref{tab:FitResults} with $1\sigma$ uncertainties for XRISM observation $201052010$.  Figure \ref{fig:CompareSpec} shows the independent fits with the above model to the different phase-filtered spectra for observation $201052010$, with a small but noticeable difference between peak frequencies of phases 3 and 1. Our phase-filtered fitting returns slightly different redshifts for the \FeKa lines from G0.11-0.11, with a slightly higher redshift when XRISM is moving away from G0.11-0.11 in phase 1 and a lower redshift when it is moving towards the target in phase 3, as expected. The phase 2 redshift is intermediate between the others, also as expected. 

Using the small-redshift approximation in Equation \ref{eq:smallred}, the redshift in terms of velocity of each phase is $v = cz$ and the difference in redshifts between phase 1 and 3 with propagated uncertainty is then $\Delta v = c \times \Delta z = (24 \pm 19) \:\rm{km/s}$ for XRISM observation $201052010$. Notably, because we subtract off the phase 3 and 1 redshift values, systematic effects that vary slowly over an observation or are uncorrelated with its orbit such as the barycentric correction or the scale uncertainty of \Resolve \citep{2024SPIE13093E..1PE} are canceled out.

The $\Delta v$ for the other observations are shown in Table \ref{tab:overallResults}, all calculated exactly identically to the illustrative example using Obs ID $201052010$.

\input{OverallResults}

\subsection{Statistical Significance of Results}
\label{sec:Stats}

While Table \ref{tab:overallResults} shows each individual observation having a positive $\Delta v$ as expected, and with the measured $\Delta v$ values being moderately close to the expected $14.7 ~\rm{km/s}$ orbital motion, it is still worth rigorously testing whether they actually represent a strong detection of orbital redshift modulation. Here we complete the hypothesis testing described in Section \ref{sec:Hypo}.

For our admittedly small sample in Table \ref{tab:overallResults}, we have sample mean $\bar{x} = 28.0$, sample variance $\frac{\sigma^2}{n} = 25.125$, and three degrees of freedom. Calculating the one-sided p-value using \texttt{scipy.stats.ttest\_1samp}, we obtain $p = 0.0084$. Because $p < \alpha$, we can reject $\mathcal{H}_0$ at the $95\%$ level. We conclude that there is evidence at the 95\% confidence level that the fitted redshift of the \FeKa lines changes in a way correlated with XRISM's motion around the Earth.

Besides this formal hypothesis test we can informally consider whether the results in Table \ref{tab:overallResults} are more likely drawn from a population with mean $\mu=0 \:\rm{km/s}$ or one with $\mu=14.7 \:\rm{km/s}$ (as opposed to $\mu>0$ of a one-sided test). Clearly, $\Delta v$ data point in Table \ref{tab:overallResults} and the sample statistics as a whole prefer $\mu = 14.7 \:\rm{km/s}$. It is worth noting that the $\Delta v$ in Table \ref{tab:overallResults} are all greater than the expected $14.7 \:\rm{km/s}$, but all are within one or two standard deviations of the expected value.

\subsection{Alternative Explanations}
\label{sec:AltEx}

Is there any other systematic effect that could imitate the redshift modulation identified above besides the motion of XRISM around the Earth?

If some cyclic effect with the same period as \textit{XRISM}'s orbit leads to a very small bias in photon energy reconstruction, a small periodic bias in line redshift might occur. For example, there is a bias in reconstructed photon energy in a calorimeter instrument like \Resolve related to the temperature of the detector \citep{2016JLTP..184..498P,2025JATIS..11d2016P,2024SPIE13093E..1PE}, with higher temperatures biasing photon reconstruction to higher energies. \textit{XRISM}'s repeated passage from shadow into sunlight once per orbit could conceivably have secondary effects on photon energy reconstruction, though this day-night cycle is not correlated with the start-stop cycle in Figure \ref{fig:LightCurveSample} that we used to extract filtered spectra, and these effects are carefully modeled in XRISM with continuous on-orbit gain monitoring, so we view these first-order effects as unlikely to produce the systematic shifts we observe in Table \ref{tab:overallResults}.

\cite{2025PASJ...77S..39M} demonstrated via observations of the Crab nebula that there is a slight offset in energy scale and resolution depending on the count rate in the \Resolve pixels. This effect could lead to variations in redshift if the count rate from an astronomical source is high and varying coincidentally with XRISM's orbit. However, because we neither expect nor detect X-ray variability from large molecular clouds in the short timescales of the observations in Table \ref{tab:overallResults}, and because the per-pixel count rate in those observations is stable and never greater than $\approx 0.1 ~\rm{ct/s/pix}$ (as shown in Figure \ref{fig:LightCurveSample} for Obs ID 201052010), we conclude that this count-rate dependent energy shift is not impacting our results.



\begin{figure}
    \centering
    \includegraphics[width=\columnwidth]{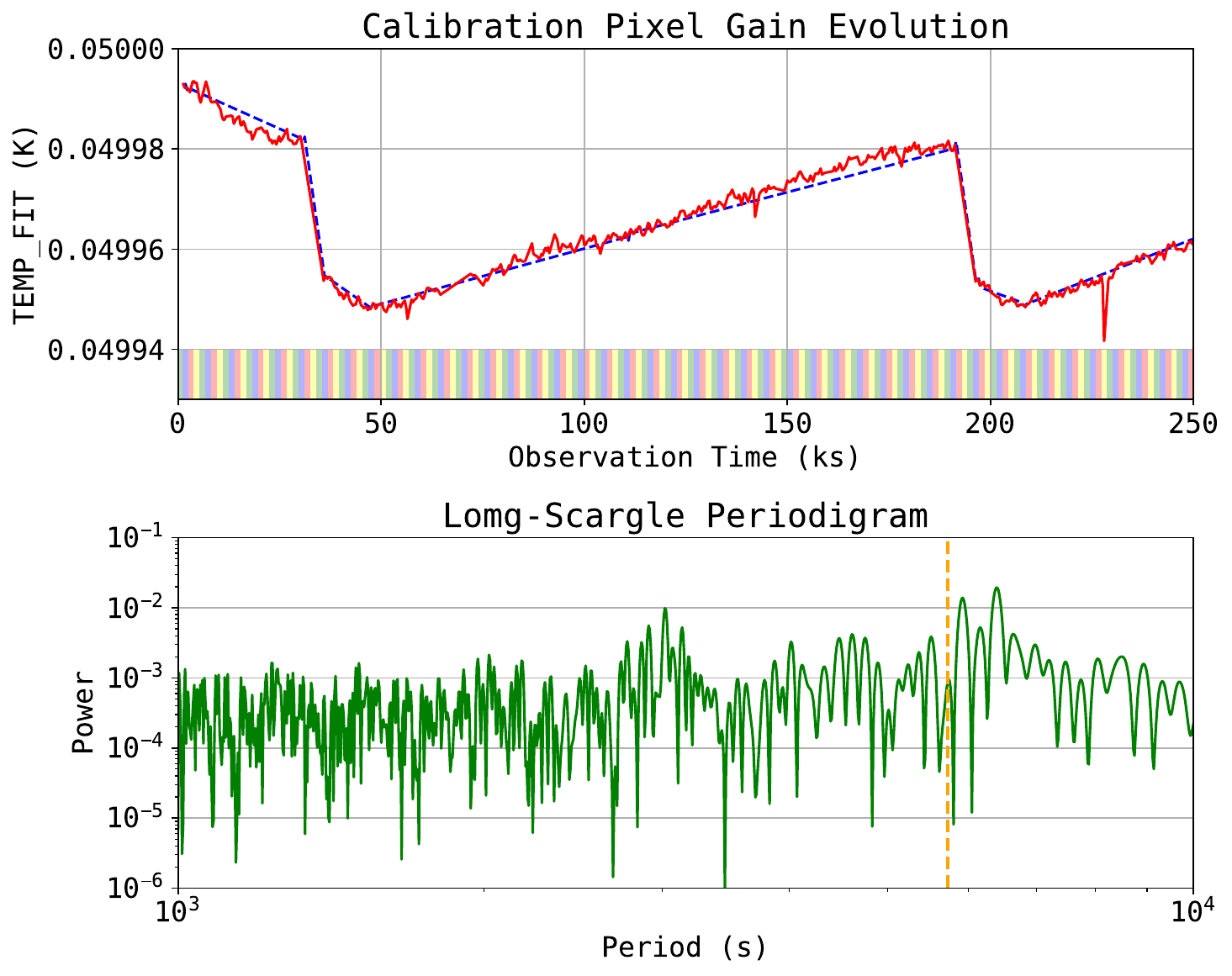}
    \caption{For ObsID 201052010, (top) Recreating Figure 9 from the gain report with the continuous (red) and intermittent (blue dashed) gain tracking of the calibration pixel, filtering out periods of detector cycling or SAA passage. Compared to the rhythm of the orbital phases (colored bars), no overall or residual trend is visible. (bottom) The Lomb-Scargle periodigram of the continuous gain monitoring in periods between $1$ and $10\:\rm{ks}$. No power excess is visible at the orbital period of $5730\:\rm{s}$ (orange dashed line).}
    \label{fig:TempFitCompare}
\end{figure}

The continuous monitoring and modeling of the gain evolution of pixel 12 -- the calibration pixel -- is an important systematic effect in \Resolve that if periodically biased could result in a modulated shift in reconstructed photon energy. To search for any unaccounted for periodic changes in this calibration, we recreate Figure 9 in the gain report for each of the observations used herein. The top panel of Figure \ref{fig:TempFitCompare} shows this calibration pixel gain history for Obs ID 201052010, with periods of detector recycling and South Atlantic Anomaly (SAA) passage filtered out. Compared with the colored phase bars, there is no obvious relationship between the gain history and the orbit of XRISM around the Sun. The bottom panel shows the Lomb-Scargle periodigram \citep{1982ApJ...263..835S,1976Ap&SS..39..447L} for the gain history at $10^4$ logarithmically-space periods between $1$ and $10 \:\rm{ks}$; no power excess at the orbital period of the satellite (or its harmonics) is evident in this or any of the other observations.

The gain reports for XRISM observations include pixel-by-pixel gain histories (Figure 10 of each report), which would show any periods of failed calibration or gain tracking that might lead to biases in energy reconstruction. If a large error in even one pixel were to occur in a short period, or repeated small errors were to accumulate periodically over multiple orbits, it might bias the overall redshift measurements. However, for the observations used herein there are no failed gain solutions for any pixels throughout the entire observations, meaning that the gain of each pixel is appropriately tracked. After consulting and confirming this with the \Resolve instrument team, we consider it unlikely that any of these biases were introduced into our measurement.

These known instrumental effects seem unlikely to produce the coherent shifts in $\Delta v$ identified by our pipeline above. As always, it is certainly possible that an ``unknown unknown'' instrumental effect might lead to the observed differences in redshift; if such an effect exists, our work here is a signal of its presence and a justification for mitigation.

\section{Discussion}
\label{sec:DiscConc}

We find evidence at the $95\%$ level for the modulation of the redshift of \FeKa lines correlated with the orbital motion of the \textit{XRISM} telescope. Lacking any alternative explanation and concluding that systematic effects in \Resolve are unlikely to produce this trend across multiple observations, the evidence suggests that we have, for the first time, observed the redshift and blueshift of X-rays photons due to the orbital motion of an X-ray telescope.

The pipeline described in Section \ref{sec:Pipe} can be applied to any X-ray spectrum to test whether the orbital redshift is a significant effect. We have produced evidence that this modulation is not just a feature of future $R>10^4$ X-ray missions on the horizon -- it is present in existing data. 

As modulation of orbital redshift is a systematic, observational effect, what could be done to mitigate it? 

As-is, the shifted energies of photons in phases 3 and 1 contribute $ 15 \cos (i) \:\rm{km/s}$ of broadening to any spectral feature, depending on the projected inclination of the satellite's orbit. Currently, this effect could be modeled with a broadening convolution during spectral analysis (not necessarily Gaussian or Lorentzian) or reduced by only using photon data from phase 2 (tangential motion across the Earth). Formally, orbital redshift modulation could be integrated by using the pipeline described above to simulate the expected broadening profile.

The energy responses of current and planned X-ray instruments like \Resolve \citep{2024SPIE13093E..1PE} and the NewAthena IFU \citep{peille2025xrayintegralfieldunit} are defined by bin widths substantially larger than the energy broadening imparted by orbital redshift. It is therefore not possible to adjust the energy of individual photons to account for orbital motion, as this shift would be smaller than the width of a single energy bin. This effect could be modeled at larger scales by redistributing photons in a particular bin to adjacent bins depending on the expected broadening profile, or by adjusting the energy axis of a response file into a geocentric frame.

In future X-ray telescopes with $R \ge 10^4$, the energy bins encoded in an extracted spectrum could be smaller than the $\Delta E$ imparted by the orbital motion of the satellite. In that case, it would be possible to adjust the reconstructed energy of individual photons during data processing to account for the orbital motion of the satellite. Existing commands to correct the arrival time of individual photons to the solar system barycenter (such as \texttt{barycen}) might be the starting point for an energy-correcting processing step. Many X-ray telescopes package spacecraft orbital information into raw data files (for XRISM, the \texttt{.orb} among others), but explicitly encoding the moment-to-moment position and velocity vectors of the spacecraft would simplify these corrections.

In our search for possible systematic biases in Section \ref{sec:AltEx}, we noted that periodic biases in calibration or pixel-by-pixel gain tracking might impact observations of narrow features or small systematic effects. The current framework of continuous gain tracking and after-the-fact quality checks used by \Resolve will be valuable as the energy resolution of future telescopes continues to increase. Though we do not expect that these effects impacted our results, we note that even finer energy resolution will necessitate even greater stability in calibration on even shorter timescales.

As far as we know, this study is the first detection of redshift and blueshift of X-ray photons due to a telescope's orbital motion. We hope it motivates discussion on how best to mitigate this effect in future high-energy missions.

\software{FTools \citep{FTools}, Xspec \citep{Xspec}, DS9 \citep{DS9}, Scipy \citep{2020SciPy-NMeth}}

\acknowledgments

Unending gratitude is due to Dr. Mihoko Yukita and other members of the \textit{XRISM} team for technical advice on the gritty details of \Resolve data processing. This work is supported by NASA XRISM cycle-1 guest observer grant number 80NSSC25K7844. No AI or LLM tools were used in the construction of this manuscript or the conduct of the science herein. We thank our anonymous referee for their insightful and useful commentary, which substantially improved this manuscript. We also thank Tobias DiKerby (age at acceptance: 4 weeks) for facilitating ruination on this topic over late-night bottle sessions.

\bibliography{main}{}

\appendix
\label{sec:Appendix}

Because the energy bin width of an unbinned \Resolve spectrum ($0.5 \:\rm{eV}$), the energy resolution of \Resolve at the \FeKa lines ($\approx 5 \:\rm{eV}$), and the bin width of the binned spectra used above and shown in Figure \ref{fig:CompareSpec} (adaptively binned but $0.5 - 1.0 \:\rm{eV}$ in the lines and $\approx 5 \:\rm{eV}$ in the continuum) are above the energy shift expected from the orbital motion of XRISM ($\approx 0.3 \:\rm{eV}$), we have no reason to expect the use of binned spectra described in section \ref{sec:XRIobs} to impact our results, especially as the binning is fine or nonexistent in the bright \FeKa lines. We conduct the following \textit{a posteriori} check on this assumption by repeating the pipeline but using unbinned spectra and the Cash statistic \citep{1979ApJ...228..939C} for fitting. 

\begin{figure}[h!]
    \centering
    \includegraphics[width=0.5\linewidth]{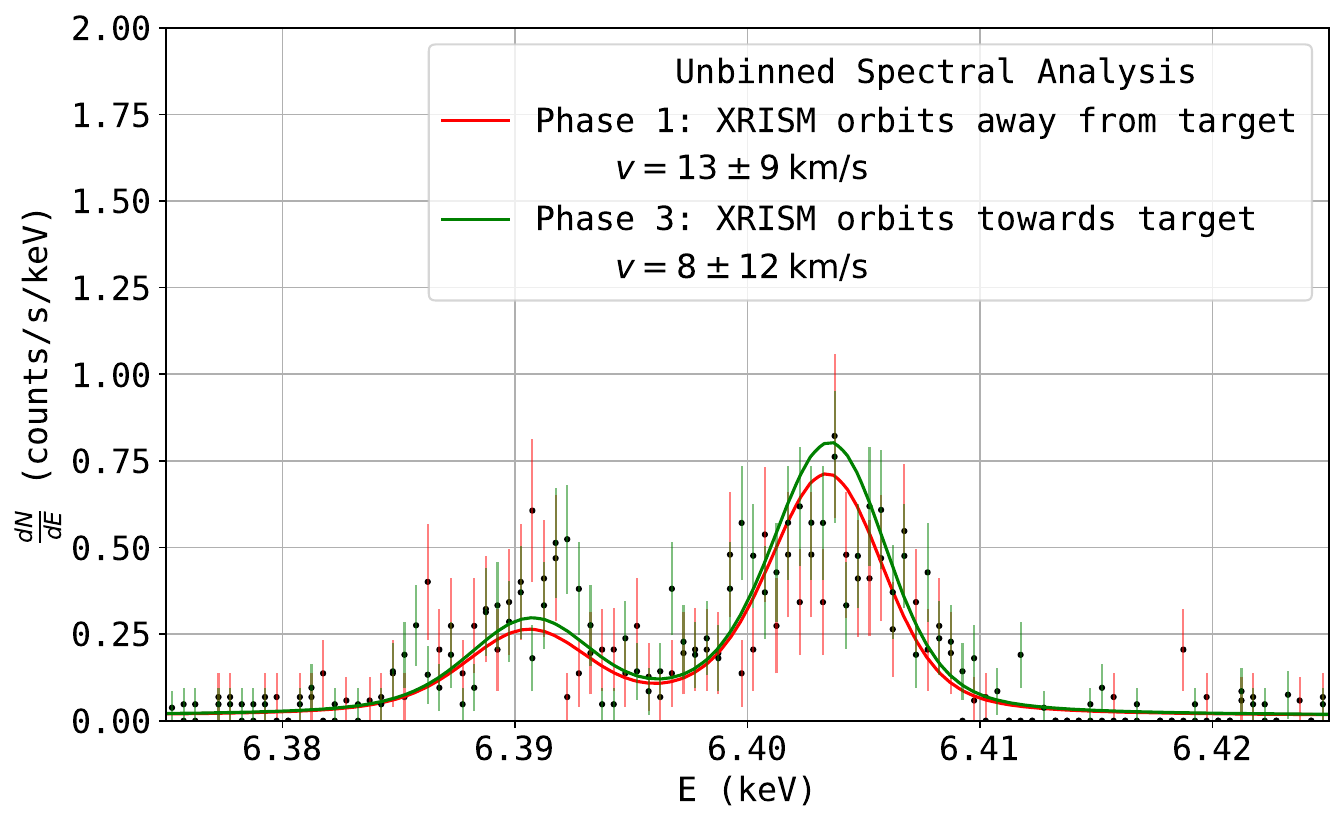}
    \caption{As in Figure \ref{fig:CompareSpec}, the phase-filtered spectra fitted with the \FeKa model from \cite{1997PhRvA..56.4554H} for Obs ID $201052010$, along with fitted redshift velocities via unbinned spectra and the C-statistic. Phase 2 has no data due to occultation of the target by the Earth, and Phase 0 is hidden to remove crowding of data points.}
    \label{fig:CompareSpec_Unbinned}
\end{figure}

Figure \ref{fig:CompareSpec_Unbinned} recreates Figure \ref{fig:CompareSpec} for ObsID $201052010$, but with the unbinned spectrum. Table \ref{tab:overallResults_Unbinned} reproduces the $\Delta v$ values in Table \ref{tab:overallResults}, showing agreement within uncertainties for each of the observations used in the pipeline above.

Using the $\Delta v$ values in Table \ref{tab:overallResults_Unbinned} as a sample for the same one-sided hypothesis test discussed in Sections \ref{sec:Hypo} and \ref{sec:Stats}, we obtain a p-value of $p = 0.037$. We should not draw any statistical conclusions from this second trial because it was conducted after the fact and motivated by the earlier result. Still, once again $p < \alpha$ and therefore we can conclude that \textit{if we had used unbinned spectra for our pipeline} the results of our hypothesis test would be the same; rejection of the null hypothesis at the $95 \%$ level.
 
\input{OverallResults_Unbinned}

\end{document}

%% file: FitResults.tex
\begin{table}
    \centering
    \begin{tabular}{lcccc}
        Phase & $N_{\rm{0,powerlaw}}$ & $N_{\rm{0,zfeklor}}$ & $z$ & $\chi^2/\rm{d.o.f}$ \\ \hline
        0 & $13.1 \pm 0.7$ & $7.1 \pm 0.4$ & $(4 \pm 3) \times 10^{-5} $ & $74.74/80$  \\
        1 & $13.2 \pm 0.8$ & $5.7 \pm 0.4$ & $(10 \pm 4) \times 10^{-5} $ & $63.07/60$  \\
        2 & & & &  \\
        3 & $14.4 \pm 0.9$ & $6.2 \pm 0.5$ & $(2 \pm 5) \times 10^{-5} $ & $63.76/52$  \\
    \end{tabular}
    \caption{Using the phases defined in \ref{fig:OrbitalMotionDiagram}, spectral fitting results for ObsID 201052010, including powerlaw ($N_{\rm{0,powerlaw}}$) and \texttt{zfeklor} ($N_{\rm{0,zfeklor}}$) normalizations in $10^{-5}\rm{/s/cm^2/keV}$, redshift, and reduced $\chi^2$.}
    \label{tab:FitResults}
\end{table}

%% file: OverallResults.tex
\begin{table}
    \centering
    \begin{tabular}{lccc}
        Target & ObsID & Exposure & $\Delta v$ \\ \hline
        G0.11-0.11 & & & \\
        & 201052010 & $120 \:\rm{ks}$ & $24 \pm 19 \:\rm{km/s}$\\
        Bridge & & & \\
        & 300045010 & $75 \:\rm{ks}$ & $24 \pm 13 \:\rm{km/s}$\\
        & 300045020 & $64 \:\rm{ks}$ & $45 \pm 17 \:\rm{km/s}$\\

        Sgr C & & & \\
        & 201046010 & $160 \:\rm{ks}$ & $19 \pm 12 \:\rm{km/s}$\\

    \end{tabular}
    \caption{Measured $\Delta v$ orbital redshifts for several XRISM Cycle 1 observations of galactic center molecular clouds with narrow \FeKa lines.}
    \label{tab:overallResults}
\end{table}

%% file: OverallResults_Unbinned.tex
\begin{table}[h!]
    \centering
    \begin{tabular}{lccc}
        Target & ObsID & Exposure & $\Delta v$ \\ \hline
        G0.11-0.11 & & & \\
        & 201052010 & $120 \:\rm{ks}$ & $7 \pm 15 \:\rm{km/s}$\\
        Bridge & & & \\
        & 300045010 & $75 \:\rm{ks}$ & $27 \pm 13 \:\rm{km/s}$\\
        & 300045020 & $64 \:\rm{ks}$ & $32 \pm 14 \:\rm{km/s}$\\
        Sgr C & & & \\
        & 201046010 & $160 \:\rm{ks}$ & $17 \pm 26 \:\rm{km/s}$\\

    \end{tabular}
    \caption{Measured $\Delta v$ orbital redshifts for several XRISM Cycle 1 observations of galactic center molecular clouds with narrow \FeKa lines, using unbinned spectra and the C-statistic for fitting.}
    \label{tab:overallResults_Unbinned}
\end{table}
